\documentclass[10pt, conference]{IEEEtran}
\IEEEoverridecommandlockouts
\usepackage{cite}
\usepackage{amsmath,amssymb,amsfonts}
\usepackage{algorithmic}
\usepackage{graphicx}
\usepackage{textcomp}
\usepackage{tabularx}
\usepackage{xspace}
\usepackage{hyperref}
\hypersetup{pdfauthor={Anonymous}, pdftitle={Paper Title}, pdfsubject={Subject}, pdfkeywords={Keywords}}

\usepackage{pifont} 

\newcommand{\ie}{\emph{i.e.,}\xspace}
\newcommand{\eg}{\emph{e.g.,}\xspace}
\newcommand{\etc}{etc.\xspace}
\newcommand{\etal}{\emph{et~al.}\xspace}

\def\BibTeX{{\rm B\kern-.05em{\sc i\kern-.025em b}\kern-.08em
    T\kern-.1667em\lower.7ex\hbox{E}\kern-.125emX}}

\usepackage{lscape}
\usepackage[table]{xcolor} 
\definecolor{lightgray}{gray}{0.9}

\begin{document}

\title{Ten Years of Software Engineering in Society}

\author{
\IEEEauthorblockN{Iffat Fatima}
\textit{Vrije Universiteit Amsterdam}\\
\IEEEauthorblockA{
Amsterdam, The Netherlands \\
i.fatima@vu.nl}
\and
\IEEEauthorblockN{Patricia Lago}
\IEEEauthorblockA{
\textit{Vrije Universiteit Amsterdam}\\
Amsterdam, The Netherlands \\
p.lago@vu.nl}
}
\maketitle

\begin{abstract}
In the international software engineering research community, the premier conference (ICSE) features since a decade a special track on the role of SE In Society (or SEIS track). 
In this work, we want to use the articles published in this track as a proxy or example of the research in this field, in terms of covered topics, trends, and gaps. Also, since SEIS was originally defined with a special focus on sustainability, we want to observe the evolution of the research in this respect.
We conducted a mapping study of the 123 articles published in the SEIS track and among the results identified (i) trends pertaining sustainability, diversity and inclusion, and open-source software; (ii) gaps regarding concrete interventions to solve problems (e.g., workplace discrimination, the emotional well-being of developers); and (iii) a main sustainability focus in the social dimension, while the environmental dimension is the least frequently addressed. As future work, our aim is to stimulate discussion in the community and we hope to inspire replications of this work in other conference venues.
\end{abstract}


\section{Introduction}
\label{s:introduction}

Technology is not just a reflection of societal needs, it actively shapes behaviors, interactions, and the way we engage with the world. As software increasingly drives user interactions and automates processes, acknowledging its socio-economic and environmental impacts becomes crucial. Analyzing these impacts on society, at large and in relation to the specific needs of underrepresented stakeholders, is therefore essential. In addition, the software engineering community itself requires support to address the diverse needs of its workforce. Ensuring inclusion and equity within industry processes is vital to foster a more balanced and supportive work environment~\cite{Albusays2021}.

The Software Engineering in Society (SEIS)\footnote{\url{http://www.wikicfp.com/cfp/servlet/event.showcfp?eventid=40910}} track of the International Conference on Software Engineering (ICSE) has played a pivotal role in exploring the impact of Software Engineering (SE) on society. It provides a platform for discussions about broad SE societal implications. This track started in 2015, welcoming research on SE for a sustainable society in various areas including health, physical-, environmental- and social sciences, management, economics, computing, policy, manufacturing, arts, and interdisciplinary research. Since 2022, the track has also welcomed a more diverse range of topics on COVID-19, ethics and diversity, and inclusion, misinformation, communication, research partnerships, and many more.

Building on past research is essential for gaining insights and advancing knowledge. Reflecting on previous work while identifying key themes and gaps helps shape future research directions and ensures continuous progress in the field. In light of this, and inspired by the fact that the SEIS track has existed for 10 years, we analyze a decade of research on the societal impact of software engineering by examining SEIS publications. We selected the SEIS track as a proxy of SE research in a societal context. From its inception, SEIS has focused on sustainability and societal impact, making it suitable for observing how the notion of sustainability evolved over time. 

To our aim, we explore trends based on the problems addressed by novel approaches and identify research gaps in areas that have received less attention. We further review the publications in this track from a sustainability perspective. We chose a sustainability angle for our analysis because sustainability is at the core of all societal values. It addresses issues related to diversity, inclusion, and equity through its social dimension, provides eco-friendly solutions through the environmental dimension, emphasizes cost-effectiveness and prosperity through the economic dimension, and enables the long-term use of digital technologies that continuously evolve to solve societal needs.

We pose three research questions (RQs) to explore this track regarding topics, trends, gaps, and sustainability foci, namely: 

\noindent\textbf{RQ1. What are the topics addressed in the SEIS track?} Through this RQ, we aim to identify the major topics published in this track. 

\noindent\textbf{RQ2. What are the research trends and gaps in the SEIS track?} Through this RQ, we aim to identify the emerging trends over a decade and research gaps that require further attention.

\noindent\textbf{RQ3. What is the coverage in the SEIS track in terms of sustainability dimensions?} Through this RQ, we aim to identify how 4D sustainability (economic, environmental, social, and technical) is addressed by SEIS publications. 

To answer these RQs, we carried out a systematic mapping study\cite{petersen2008systematic} to map the state of the research published in SEIS. 

The rest of the study is structured as follows. Section \ref{s:bg} provides the study background. Section \ref{s:related} presents other similar publications that map state-of-the-art in SE publications for other conferences. Sections \ref{s:method} and \ref{s:results} describe the study design, and the study execution and results, respectively. Section \ref{s:discussion} discusses our reflection on the results. Section \ref{s:threats} provides an overview of the threats to the validity of this research and mitigation strategies. Finally, Section \ref{s:conclusion} concludes along with some future directions.

\section{On Sustainability}
\label{s:bg}

As we use a sustainability perspective for our analysis, here we provide context for sustainability in terms of its need and impacts, followed by a description of sustainability-related concepts that are used in the analysis.

\subsection{Background}
The desire to incorporate sustainability into SE stems from society's growing understanding of sustainability. Products that are not only effective and user-friendly, but also ethical and environmentally responsible, are becoming increasingly popular\cite{mainieri1997green}. Approximately 97\% of climate experts believe that human activity is mostly responsible for the trends in global warming over the last century\cite{cook2016consensus}. In today's digital age, software systems are critical to the functioning of civilization, influencing everything from complex industrial operations to daily communication \cite{krohn2018software}. As a result, the development, deployment, and maintenance of these systems have significant ramifications that extend beyond conventional metrics like speed and reliability to include long-term effects on the environment and society \cite{penzenstadler2014sustainability}. 

Despite these advances, there are numerous challenges and knowledge gaps in the field of SE for sustainability\cite{venters2017characterising}. As a result, sustainability and societal considerations are often considered secondary, pointing to a fundamental barrier in changing SE to more responsible and future-oriented approaches. There are notable variances in how sustainability is implemented across different industries and regions, which can be attributed to varying levels of resources and expertise \cite{noman2022exploratory}. Furthermore, some parts of sustainable software engineering, such as energy-efficient computing, have received more attention than others, such as the social ramifications of software systems and their role in long-term economic stability. Our study seeks to critically examine the current state of sustainability-related research in the SEIS track. 

\subsection{Definition, dimensions, and impacts}
In the context of this study, we define sustainability as ``the preservation of the long-term and beneficial use of digital solutions, and their appropriate evolution, in a context that continuously changes’'~\cite{sust-def}. This definition implies that an equitable way of problem-solving must be established and that the impacts of SE solutions must be carefully analyzed over time. Moreover, to classify the types of ``beneficial uses'' of digital solutions in the specific context of SE, we use the four sustainability dimensions defined by Lago \etal~\cite{lago2015framing}. In particular:

\textit{Economic dimension} refers to preserving capital and financial value. 

\textit{Environmental dimension} refers to the preservation of natural resources by addressing ecological requirements.

\textit{Social dimension} refers to the preservation of social resources through generational equity by supporting and creating benefits for communities. 

\textit{Technical dimension} refers to the preservation of software in terms of its long-term use and continuous evolution.

In addition, to analyze the role of SE in society, we consider both its primary and its enabled focus in studies based on direct and enabling effects defined by Hilty \etal~\cite{hilty2011sustainability}.

\section{Related Work}
\label{s:related}

Works related to ours include publications that review the research published in SE scientific venues, in general, as well as in the context of sustainability. 

\subsection{Conference Track Reviews}
Few works review SE conference tracks to capture the socio-technical perspectives and human values, described as follows.

A classification of ICSE publications (research track) from 2015-2017, captures the socio-technical perspectives ~\cite{williams2019methodology}. The study identifies a need to diversify the research techniques to include human and social aspects while maintaining a balance with the technical aspects. The results show that stakeholder involvement using design science strategies, and using human subjects for research while fulfilling all ethical criteria, can aid in including socio-technical aspects. Triangulation and diversification of research strategies can also help improve this.
An investigation of human values in SE publications (from 2015-2018)~\cite{Perera2020} identifies 11 categories of human values from ICSE, ESEC/FSE, TSE, and TOSEM, based on the Schwartz Values Structure. The results of the study reveal that only 16\% of the publications considered human values, with 41\% related to security. Furthermore, the findings show that 60\% of the socially significant values are ignored in SE research.

\subsection{Sustainability research in SE}
Several systematic mapping studies have been performed over the years to observe the state of sustainability research in SE and its evolution over time.

A systematic mapping analysis was used to classify sustainability research within SE~\cite{penzenstadler2014systematic}. The study identified various research hotspots, including models, techniques, and software design, and mapped publications to knowledge domains. The report offers a detailed summary of current trends in sustainability research and suggests avenues for further investigation. The study mapped the broader subject of sustainability in SE. Another mapping study further classifies state-of-the-art concepts, models, and frameworks in relation to sustainability in SE~\cite{mourao2018green}. Penzenstadler \etal~\cite{penzenstadler2012sustainability} carried out a comprehensive literature review to understand the current status of sustainability research in SE. They gave an overview of the body of existing literature and classified research efforts into various sustainability issues. A multi-vocal literature review assessing the practical relevance of SE research over 34 years (1985-2019) \cite{Garousi2020PracticalRelevance} identifies a lack of relevance and collaboration with the industry. The study also emphasizes the significance of carrying out empirically grounded research on the notion of relevance in SE. 

Previous studies have examined the state of research in either (i) general software engineering tracks~\cite{williams2019methodology,Perera2020} or (ii) studying one aspect \eg sustainability~\cite{penzenstadler2012sustainability, penzenstadler2014systematic,mourao2018green} or practical relevance~\cite{Garousi2020PracticalRelevance} over the years. Contrary to the previous works, we analyze the SEIS track, which is dedicated to SE in society research. Through this study, we aim to identify the topics and emerging trends in this domain and the impact of this track in terms of its research contribution. 

\section{Study Design}
\label{s:method}

To conduct our systematic mapping study, we use the methodology described by Peterson \etal~\cite{petersen2008systematic}. We provide a replication package~\cite{rep-pck-icse-seis-2024} that contains an appendix with primary studies, extracted data, and scripts. 

\subsection{Data Collection}
We collected the publications published in the ICSE SEIS conference track from its inception, 2015-2024. We retrieved a total of 123 publications. We provide the list of publications with their publication IDs as an Appendix in our replication package~\cite{rep-pck-icse-seis-2024}.

Contrary to the typical inclusion and exclusion criteria, our study does not use this step from Peterson \etal~\cite{petersen2008systematic}. As our study requires an analysis of all publications published in this track, we include all publications for data extraction. 

\subsection{Classification Scheme and Mapping}
Based on the goals of our RQs, we classified and mapped the publications to relevant categories as follows. 

\subsubsection{RQ1 -- Research Topics}
We extracted the keywords from the respective study's metadata. For the publications with missing keywords, we used the abstracts as an extraction source. If the abstract did not provide sufficient information, we used the introduction and conclusion of the publication as a secondary keyword source.

We collected a total of 644 keywords from the 123 publications. We clustered these keywords to observe the frequency of research topics. We used BERT model~\cite{BERTdevlin2019} to generate embeddings that were used to create semantically similar clusters using K-means clustering. The scripts are provided in our replication package~\cite{rep-pck-icse-seis-2024}. We excluded certain keyword clusters from thematic categorization. These include keywords like SE, Software Development, Technology (and technology names). This exclusion was performed as these keywords did not represent the research theme of the study under analysis. As all publications are published at ICSE, the high frequency of such keywords is a natural consequence. Hence, we exclude them. Further, we manually merged keyword clusters based on overlapping themes. 

\subsubsection{RQ2 -- Types, Trends and Gaps}
We categorize the publications based on the type of research using the classification by  Wieringa \etal~\cite{wieringa2006requirements}. These categories include evaluation research, validation research, proposal of solution, philosophical publications, personal experience publications, and opinion publications. We aim to understand the research methodologies employed to study SE in society. We classify the publications against a research type by initially reviewing the introduction and conclusion sections. Further, we used key phrases and approaches described in the full text to assign the publications to relevant categories. For example, phrases like `evaluation of', `case study', and `real-world application' frequently indicated Evaluation Research, whereas phrases like `proposes a new method' and `introduces a technique' suggested a Solution Proposal. These indications were used to hypothesize a research type. The final decision about the research type was made based on the type definitions by Wieringa \etal~\cite{wieringa2006requirements}.

Based on the keyword clusters, we classified the publications by research focus and contributions, identifying trends based on the distribution of publications over time. Emerging trends were labeled with descriptive names, while categories with fewer publications revealed research gaps. These gaps were highlighted to guide future research. We organized the categories of trends based on the keywords used in the papers. For instance, a paper on diversity and inclusion may fall under sustainability, but we only included it if it was framed as sustainability in the paper. We read the full text of these publications to confirm their categorization into a trend and reported on their contributions. 

\subsubsection{RQ3 -- Sustainability Coverage}
To answer RQ3, we read the publications with a sustainability lens. We provide for each publication, two types of sustainability mappings: (i) Primary sustainability focus that for each publication maps the primary focus of its contribution to the corresponding, single sustainability dimension; and (ii) Enabled sustainability focus that for each publication maps the intent of its contribution on one or multiple corresponding sustainability dimensions. For example, a publication contributing design patterns for inclusive human-computer interaction would be classified as (i) a technical contribution (for software design) with (ii) a social intent (for user inclusivity).

\subsubsection{Limitations}
As suggested by Verdecchia \etal~\cite{verdecchia2023}, here we discuss some limitations posed by our study design. Section~\ref{s:threats} will discuss the threats to the validity of our empirical investigation. For RQ1, as we mainly used the keywords provided by the publication authors, our interpretation of the topics might be limited. To avoid the propagation of this possible bias in RQ2 results, we reclustered the keywords by manual analysis and merged them into different clusters. Further, we read the publications' text in full to extract the trends based on the context of the publications. Based on the full-text check, none of the papers were removed from the clusters, rather only merged into larger generic categories. This shows that the original keyword clusters were adequate for topic classification. For RQ3, the sustainability classification is subject to the authors' understanding of sustainability. To mitigate possible limitations or biases, different co-authors performed the two types of sustainability classifications and cross-checked the results for consensus. Finally, by design the scope of our study is limited to the SEIS publications. As such, all our results are not generalizable. However, future research could extend this study with a broader scope, \eg to map the general state-of-the-art about the role of SE in society.

\section{Study Execution and Results} 
\label{s:results}
In this section, we present our results organized per RQ.

\subsection*{\textbf{RQ1. What are the topics addressed in the SEIS track?}}
\begin{figure*}[ht!]
    \centering
    \includegraphics[width=0.75\linewidth]{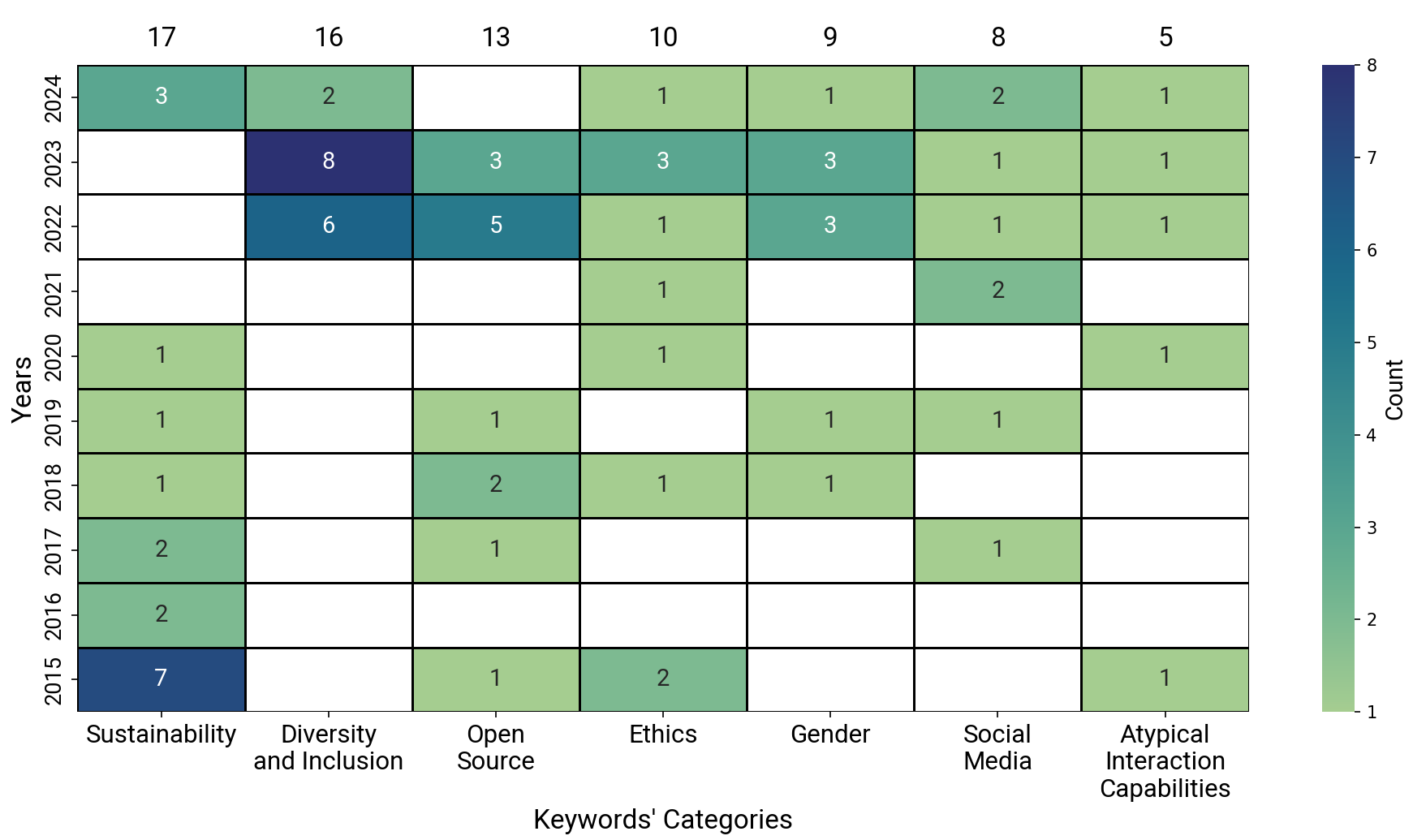} 
    \caption{Top Keywords Categories - Frequency over the years}
    \label{fig:top-trends}%
\end{figure*}

We identify the top seven\footnote{We report on the top 7 only. The rest of the categories have a relatively low frequency. Data is provided in the replication package~\cite{rep-pck-icse-seis-2024}} categories from the keyword clusters (see Fig.~\ref{fig:top-trends}), as detailed in the following. 

Sustainability appears as the top keyword in this track, used by 17 publications. Next, diversity and inclusion are the top keywords used by 16 publications. A total of 14 publications mention keywords specific to their research methodology, where case publications and empirical research are the commonly used keywords. Thirteen publications discuss open-source software. Ten publications discuss ethics with topics ranging from concerns, micro-politics, policy, and whistleblowing phenomena to the use of harmful terminology and the responsibility associated with such ethical issues. Nine publications include topics on gender bias, gender balance, and inequities. Eight publications collect and analyze data from social media platforms such as Twitter/X, Reddit, LinkedIn, Facebook, and StackOverflow.
Seven publications focus on Education, both enabled through software and the education of software engineers. Five publications focus on solving problems for individuals with atypical interaction capabilities such as dementia, autism, visual impairments, \etc

From Fig.~\ref{fig:top-trends}, we can observe that:
\begin{itemize}
    \item Sustainability was a primary focus in the early years, followed by a noticeable gap from 2021 to 2023, and a renewed interest in 2024. This may be a reflection of the initial focus of the track (which in the 2015 call featured sustainability prominently), and the recent increase in sustainability-related societal and research focus.
    \item Discussions surrounding open source were inconsistent, peaking in 2022.
    \item Topics related to ethics and gender have been present since 2018; however, publications explicitly addressing diversity and inclusion began to emerge only after 2022, albeit more frequently.
    \item From 2016 to 2019, and in 2021, there is a notable gap in research about the atypical interaction capabilities of users and developers. From 2022, however, this topic seems to be tackled again stably even if with just one publication per year. 
\end{itemize}

\subsection*{\textbf{RQ2. What are the research trends and gaps in the SEIS track?}}
Based on the main keywords, we classify the related articles into various topics. 

In the following, the trends reflect the topics that are most frequently addressed, while the gaps reflect the topics addressed by fewer publications and limitations of the published work. In the next subsections, we discuss the trends and gaps in terms of research types, and problems explored, and solutions provided.

\subsection{\textbf{Trends}}
\begin{figure}[!ht]
    \centering
    \includegraphics[width=\linewidth]{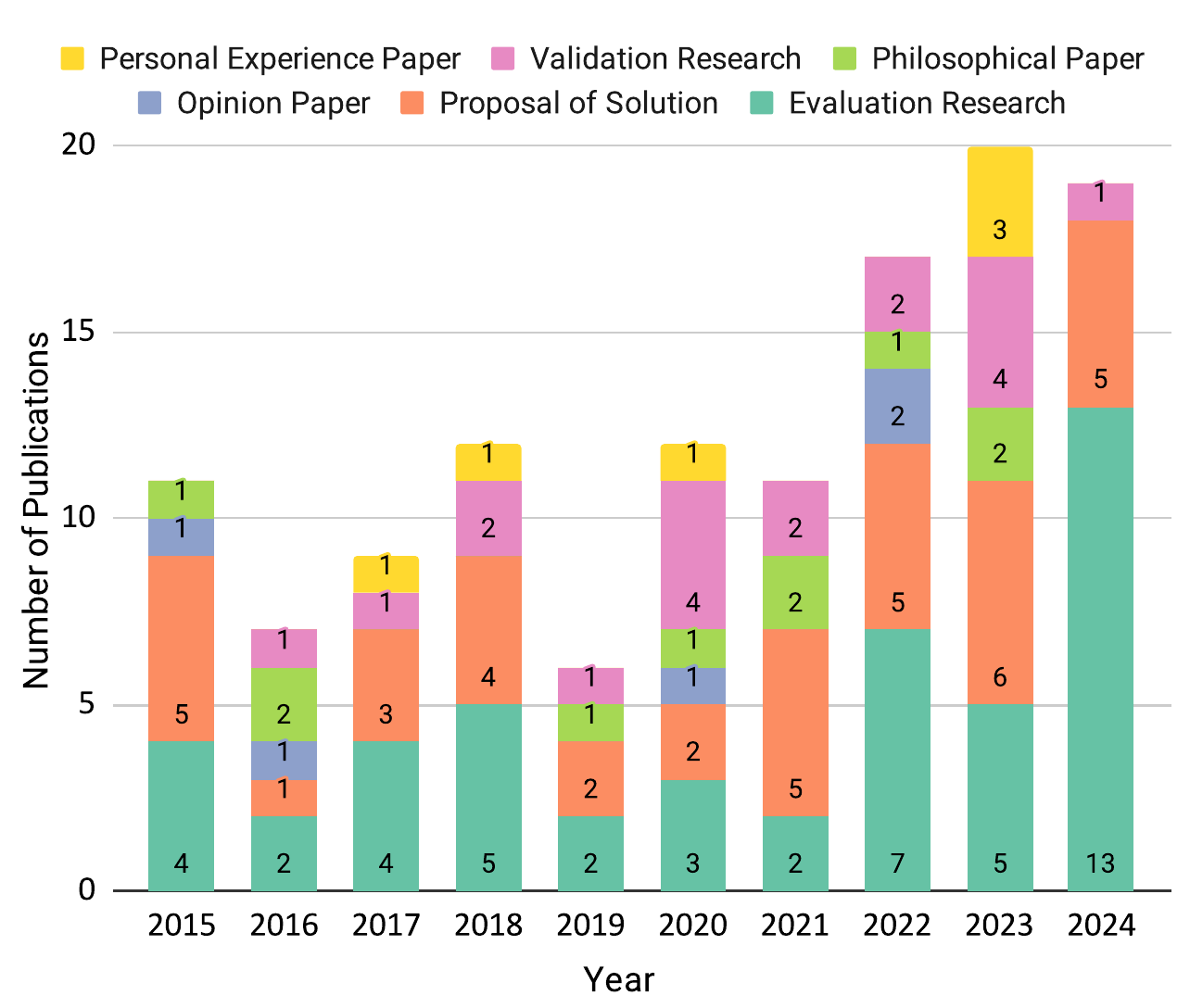} 
    \caption{Research Methodology Types over the years}
    \label{fig:research-types}
\end{figure}

\textbf{Research Types.}
Fig. \ref{fig:research-types} shows a general increase in the number of publications published annually, indicating a rise in interest and research. There is a noticeable emphasis on empirical research and rigorous scientific procedures, as evidenced by the frequent attention given to evaluation research methodologies. Research on evaluation and solution-based approaches peaks from 2022 onwards. Solution-based publications are distributed unevenly through the years, maintaining a steady number from 2021 onwards. The less common philosophical publications constantly contribute theoretical results; personal experience and opinion articles, on the other hand, are more common yet provide deep practical insights and subjective opinions.

\textbf{Sustainability.}
Research in this category highlights the multi-dimensional nature of sustainability in software engineering, spanning design (technical), human needs (social), ethical considerations (social), and collaborative strategies (socio-technical), all of which contribute to building sustainable software.
Our results show publications discussing sustainability in terms of general principles around sustainability design [P01], special human needs in the context of sustainability [P10, P11], sustainability design through requirements [P13, P19] and architecture [P14, P45, P107, P118], and socio-technical sustainability [P54]. Other topics include the identification of ethical [P09] and value-sensitive concerns [P04], sustainability assessment criteria [P116], sustainability through language [P05] and governance [P06], collaboration strategies for sustainable SE [P21], and inter-disciplinary sustainability initiatives [P36]. 

\textbf{Gender Inequity.} Multiple publications highlight gender inequities in SE faced by women and their contributions to SE roles. Bias in job advertisements discourages female candidates [P69], and women encounter issues such as cultural sexism, work-life balance difficulties, imposter syndrome, and the glass ceiling [P71]. Despite these challenges, women play a significant role in reducing miscommunication and information overload in SE teams, even when outnumbered [P40]. There are systemic challenges that women face in their career trajectory from university to workplace which require support at all stages [P120]. 
    
In terms of gender diversity, LGBTQIA+ professionals face unique challenges in remote work [P90] while also bringing unique strengths to the job, while SE professionals with immigration backgrounds experience various forms of micro-inequities [P108]. Research on female open-source contributors points to the competence-confidence gap [P38], though a general increase in their participation has been observed, with a slight drop during the COVID-19 pandemic [P85].
    
Additionally, one study explores men’s attitudes toward gender equality at the workplace, identifying both supportive and hindering behaviors [P88], while children’s perceptions show a more balanced view of SE roles for men and women, with the pandemic further normalizing certain roles due to increased accessibility [P89].

\textbf{Adapting for individuals with atypical interaction capabilities.}
Publications in this category focus on user interface adaptations for the needs of users with atypical interaction capabilities. Trends show publications focusing on (i) challenges in engineering software; for dementia [P10], designing UI for autistic users by preventing frustration and mental exertion caused by animations [P78], and identifying the human-centric issues on GitHub for visually impaired and dyslexic users [P76], and (ii) leveraging software for inclusivity; by integrating social interaction and visual assistance during physical activity [P50], and using an inclusive design process for providing navigational assistance to users with cognitive impairments [P95].


\textbf{Ethical concerns in SE.}   
We see a trend in publications exploring ethical considerations and social issues in software development and technology use. One study argues that ethical concerns are too complex to be governed by rigid rules or treated as simple non-functional requirements [P09]. To address these complexities, a technique for generating augmented regulatory text has been developed to aid both developers and policymakers in promoting principled morality [P35]. Another study investigates the failure of an ERP system, identifying micropolitical intervention as a primary factor [P53]. Privacy comparisons between different types of apps reveal that COVID-19 apps tend to manage privacy and ethical issues better than social media and productivity apps [P63]. The experiences of marginalized communities on social platforms are also highlighted, with common concerns including discrimination and misrepresentation [P92]. A tool has been created to detect and replace harmful terminology, promoting inclusivity across race, gender, ability, and neurodivergence [P97]. Recommendations for improving whistleblowing practices in SE are provided, focusing on harm mitigation and the role of professional bodies [P70]. Surveys show that many respondents view social awareness and ethics as significant concerns for smart devices used in public spaces [P101]. Additionally, using humor has been found to improve developer engagement, particularly in challenging tasks like testing and documentation [P106]. These findings underline the importance of integrating ethical principles throughout the use and development of software, from privacy and inclusivity to social awareness and organizational responsibility.

\textbf{Open Source Software (OSS).} 
We identified several publications discussing various aspects of open-source software (OSS) like governance, inclusivity, and contributor dynamics. Governance rules are proposed to enhance transparency, traceability, and semi-automation within OSS projects [P06]. Efforts like the Software Heritage Archive aim to collect, preserve, and make source code publicly accessible [P28]. The influence of large foundations on OSS development is also examined [P29]. Further, a study investigates the motivations behind projects joining the Apache Software Foundation. These motivations stem from community-building, project strengthening, and enhanced technical development [P84]. The impact of perceived gender identity and code quality on pull request acceptance decisions is analyzed, highlighting how these factors shape contribution evaluations [P44]. A dashboard designed to attract and retain OSS contributors is presented [P72], alongside an inclusivity debugging process that addresses information architecture faults [P77]. Surveys of OSS contributors reveal ongoing challenges related to diversity and inclusion, focusing on gender, seniority, and language proficiency [P81].  Research on gender differences in code contributions indicates a positive trend in women's participation, though there was a slight decline during the COVID-19 pandemic [P85]. Additionally, the tendency for women to withdraw earlier from OSS participation compared to men is noted [P104]. Finally, it is argued that insights from scientific research with social impact should be treated as open-source software artifacts to maximize their reach and utility [P99].

\textbf{Social Media Analysis.}
Our findings reveal a growing trend of utilizing social media platforms to mine valuable data for understanding user needs, ethical concerns, and developer perspectives.
Several publications use social media for mining user needs, such as mapping Twitter/X data to enhance emergency app features [P25], analyzing images from Twitter/X during the COVID-19 pandemic [P63], and scraping subreddit data to identify ethical concerns of marginalized communities [P92]. Social media is also used to extract developer perspectives, like analyzing Stack Overflow conversations to understand secure coding practices [P42]. Additionally, online discussions are analyzed, including app store reviews of COVID-19 apps for security and accessibility [P68] and LinkedIn discussions to improve scientific communication [114].

\textbf{Citations Trends.} To further elaborate on the trend and impact of research, we rank the publications by citation count. We use the  \textit{crossref} API\footnote{Retrieved on 14-02-2025 via CrossRef \url{https://www.crossref.org/documentation/retrieve-metadata/rest-api/}} to extract citation data. One limitation of citation analysis is that such APIs may provide partial citation data, as the availability of citations is dependent on the completeness of citation meta-data\footnote{Actual citation numbers may vary. \url{https://community.crossref.org/t/missing-article-citations/5631}}. 

\begin{figure}[!htbp]
    \centering
    \includegraphics[width=\linewidth]{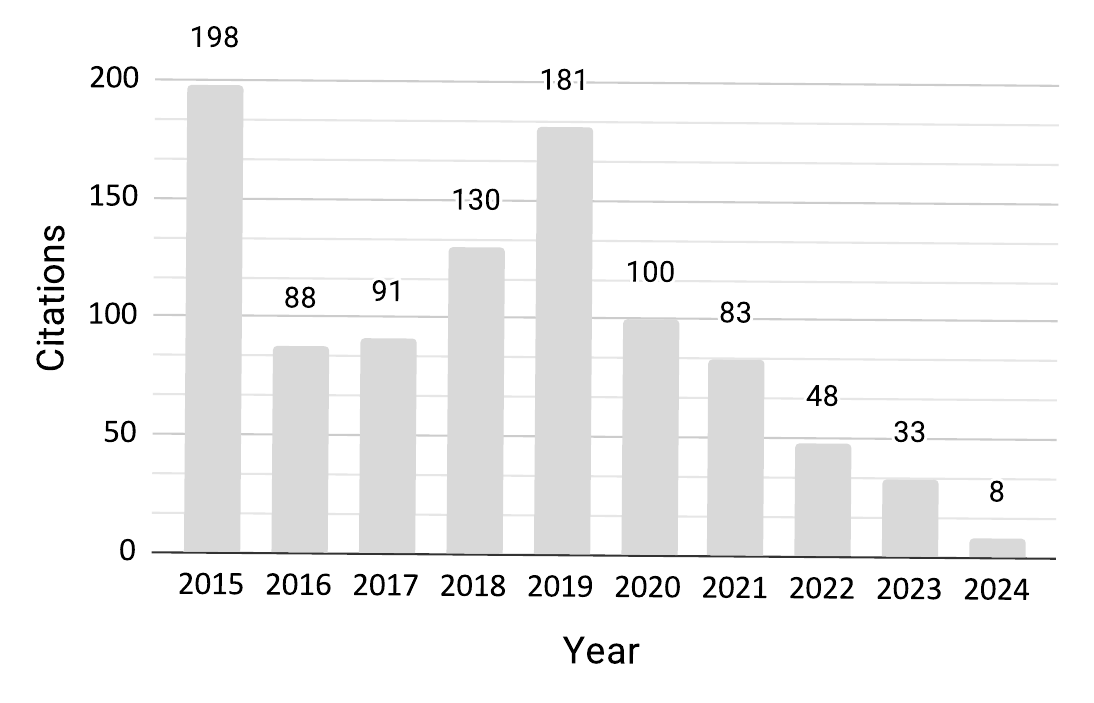}
    \caption{Citations of SEIS papers over the years}
    \label{fig:citations}
\end{figure}

Naturally, older papers tend to have more citations (see Fig.~\ref{fig:citations}). Many post-2020 papers have 0-5 citations. However, some topics may attract varying attention over the years. For top trends, we focus on papers with at least 10 citations. The most frequently cited research pertains to sustainability, with the Karlskrona Manifesto [P1] leading the list. Other notable sustainability papers discuss decision-making [P45], sustainability requirements [P19], sustainability debt [P14], and value-sensitive design in SE [P04]. After sustainability, the topic of gender diversity, specifically women in technology teams [P40], follows with the second highest citations. Additional papers concerning the role of women, and related challenges and impacts, also have substantial citations [P45, P19, P14, P04, P71]. Numerous citations are recorded for studies about value-driven SE [P15, P57, P61, P04]. Moreover, there is a high citation volume for research related to App Stores and Social Media. We provide the citation data in our replication package~\cite{rep-pck-icse-seis-2024}. We cannot infer much from these citation numbers, the only observation we make is that papers with a wider scope have a high citation count. Other papers, which might otherwise be presenting significant work in a niche area may have fewer citations. 

\subsection{\textbf{Research Gaps}}

\textbf{Research Types.}
Fig.~\ref{fig:research-types} shows a small number of opinion publications and personal experience publications. We observe that practical, experience-based contributions and subjective interpretations are lacking. This gap identifies possible areas for additional research and involvement, where the discipline could benefit from a greater focus on real-world experiences and opinion-based conversations with experts and industry.

\textbf{SE for Low Socio-Economic Groups.}
Despite growing interest in designing software for diverse user groups, limited research specifically addresses the needs and challenges of low socioeconomic groups [P94]. Factors like low literacy rates and socioeconomic status significantly impact how these groups use software. More research is needed to develop software that is accessible and usable for these user groups.

\textbf{Role of Emotions in Software Development.}
While the technical aspects of software development are well-studied, there is limited research on how developers' emotional health affects their job satisfaction and productivity. In addition to documentation [P67], understanding developers' attitudes toward SE processes could improve SE practices and enhance their well-being and productivity. Future research should explore these emotional aspects to better support SE practice.

\textbf{Impact of Workplace Discrimination Interventions.}
While the trends show publications discussing the prevalence and consequences of workplace discrimination, little research has been done to examine the efficacy of interventions to reduce and mitigate discrimination. Further research is needed to evaluate strategies and policies in this area [P105].

\textbf{Multidisciplinary Approaches for Smart Public Spaces.}
Designing user-friendly smart public spaces requires a multi-disciplinary approach, yet there is little research in this area. Future publications could integrate insights from SE, social sciences, and urban planning to tackle the challenges and opportunities of smart public spaces, making smart city initiatives more inclusive and effective [P101].

\textbf{Social Dynamics in SE Teams.}
Despite extensive research on the technical aspects of SE, the social dynamics within SE teams have received little attention. Understanding interpersonal relationships, communication styles, and team dynamics is crucial for enhancing collaboration and productivity. While existing literature highlights the importance of societal factors like socioeconomic issues and workplace discrimination, a gap remains in understanding how social dynamics affect project outcomes and individual well-being. Future publications should explore how social interactions and team cohesion influence software development processes and results [P67, P105, P94].

\subsection*{\textbf{RQ3.What is the coverage in the SEIS track in terms
of sustainability dimensions?}}

\begin{figure}[!b]
\centering
\includegraphics[width=\columnwidth]{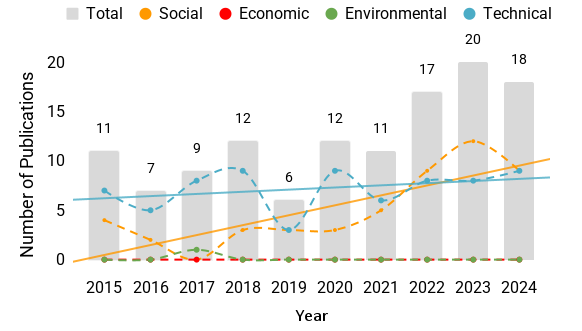} 
\caption{Primary Sustainability Focus over the years}
\label{fig:sus-primary}
\end{figure}
\begin{figure}[!b]
\centering
\includegraphics[width=\columnwidth]{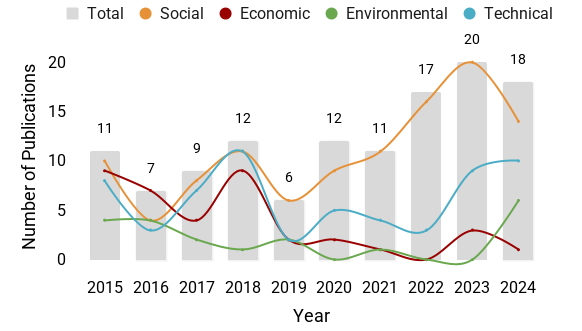} 
\caption{Enabled Sustainability Focus over the years}
\label{fig:sus-enabled}
\end{figure}
We classify the publications based on their primary and enabled focus on some sustainability dimensions. Figs.~\ref{fig:sus-primary} and \ref{fig:sus-enabled} illustrate the yearly distribution of SEIS publications on sustainability dimensions. Each bar represents the total publications published in a year. The lines and trend lines represent the coverage of the sustainability dimensions of the focus, \ie, social, economic, environmental, and technical.

\textbf{Primary Sustainability Focus}
In Fig.~\ref{fig:sus-primary}, we use the sustainability dimensions of focus to classify the ``direct'' novel contributions of SEIS publications. 
We map all publications to one of the four sustainability dimensions in terms of their primary focus. Our results show that most publications (72 out of 123) are technical (including support for SE processes, and proposal of software applications, tools, and approaches). A significant number of publications (50 out of 123) focus on social aspects (including analyses of and reflections on social issues, cultures, hiring procedures, skills, and gender issues). Only one publication [P27] contributes to environmental aspects (environmental awareness creation), while no publication directly has an economic type of contribution. 

Fig.~\ref{fig:sus-primary} also shows the trend line of the primary contribution of publications in the social and technical dimensions. The two trend lines show a slight (for the technical contributions) and a steeper (for the social contributions) increase in publications. Interestingly, they intersect in 2022 when the focus on social aspects surpasses for the first time the focus on technical ones.

\textbf{Enabled Sustainability Focus.}
We use the sustainability dimensions of focus to classify the ``enabling'' impact of the contributions to one or multiple sustainability dimensions. Fig.~\ref{fig:sus-enabled} shows the trend of the overall sustainability coverage in terms of both direct and enabling impacts. Contrary to trends based on direct contribution, we observe that the social dimension is the most addressed (109 out of 123), reaching its peak in 2023. The technical dimension is the second most addressed dimension (62 out of 123). However, even in terms of enabling impacts, the economic dimension is less emphasized (38 out of 123), and finally, the environmental dimension is the least discussed (20 out of 123).

We present a summary of the sustainability classification in Table \ref{tab:sustainability}, showing contributions per year across all sustainability dimensions. We summarize the overall sustainability trends as follows. 

\textbf{Economic dimension.} Publications in this area highlight cost efficiency, resource management, and financial benefits from sustainable practices [P01-P10, P13-P18, P27, P41, P45, P99]. However, interest in the economic dimension has declined in recent years, with few contributions in 2023-2024 [P116].

\textbf{Environmental dimension.} Early publications prioritize energy efficiency, renewable energy, and reducing environmental footprints [P01, P03, P05, P13, P19, P27]. After a dip in attention between 2019 and 2023, interest resurges in 2024 with a focus on energy consumption and sustainability-aware architecture [P107, P110, P118, P119].

\textbf{Social dimension.} This is a consistent area of interest, with an emphasis on human-centered design, and inclusivity [P01, P04, P10, P15, P32, P38]. Recent publications address equity, diversity, and support for marginalized groups [P40, P76, P82, P89, P120], showing an evolving focus on social equity.

\textbf{Technical dimension.} Throughout the years, there has been a strong focus on maintainability, adaptability, and long-term usability of software [P19, P22, P28, P33, P99]. Recent publications include topics like privacy, security, and architectural patterns [P109, P112, P118].

\begin{table*}[!htbp]
\centering
\caption{Overview of papers per publication year: classification per sustainability dimension of focus
}
\rowcolors{1}{white}{lightgray}
\begin{tabular}{|p{0.05\columnwidth}|p{0.35\columnwidth}|p{0.35\columnwidth}|p{0.5\columnwidth}|p{0.5\columnwidth}|}
 \hline
{\bf Year} & {\bf Economic Focus} & {\bf Environmental Focus} & {\bf Social Focus} & {\bf Technical Focus}\\
\hline
2015 
& 
Publications focus on cost efficiency [P01, P02, P03, P04, P05, P06, P07, P09] and savings through efficient resource use [P10]
& 
The biggest environmental trend was energy efficiency [P01, P05]. Publications focus on renewable energy sources [P03] and reducing energy use [P01, P05]. 

& 
Publications focus on community engagement [P01, P03, P06, P07], support for social issues [P10, P09],  human values [P04], encouraging social connection [P01], supporting social activity [P10], and developing inclusive and socially aware computer systems [P10, P08]. 
&
Publications focus on scalability [P03, P08], adaptability [P06, P10], reuse [P02], maintainability and flexibility [P06], evolution [P05], energy-efficient designs [P01, P05], long-term use and adaptation [P01, P05, P10], and robust and resilient software systems [P01].
\\
2016 
& 
Publications focus on cost efficiency [P13, P14, P15], resource management [P12, P17], market competition [P16], and economic prosperity~[P18].
& 
Publications focus on energy efficiency [P12, P14], the use of renewable energy [P13], and energy policy [P18].
&
Publications focus on the social benefits of architectural strategies [P14], human values [P15], community engagement [P17], and health [P18].
&
Publications focus on maintainability [P14, P15], adaptability [P15], and integration and flexibility [P17]. 
\\
2017 &
Publications focus on the financial benefits of sustainable software methods, focusing on large cost savings and enhanced economic outcomes [P27, P19, P23, P24].
&
Publications focus on new design and engineering approaches [P19, P27] for reducing the energy footprint of software systems.
&
Publications focus on community building and trust [P24, P25, P26, P22]. The research also focused on the societal impact of SE [P20, P19], improving community health [P23] and sustainable community [P21].
&
Publications mostly focus on maintainability [P19, P20, P21, P22, P24], followed by continuous analysis and integration [P25, P27].
\\
2018 
&
Publications focus on cost efficiency [P32, P34, P35, P36], cost savings [P32, P33, P34, P35, P36] and financial sustainability~[P38].
& 
Resource efficiency [P36] was the only topic discussed in environmental sustainability. 
& 
Publications focus on accessibility [P36], human-centered design [P31, P32], inclusion [P30, P39] and gender diversity [P38].
& 
Publications focus on adaptability [P28, P30, P31, P32, P36, P37], maintainability [P29, P31P33, P34], and scalability and flexibility [P39].
\\
2019 
&
Publications focused on economic value [P41, P45], market competitiveness [P41], and lowering operating expenses [P45].
& 
Publications focus on SE's role in water management [P45] and environmental decision-making [P45]. 
& 
Publications focus on gender [P40], social factors in code contributions [P44], and community engagement[P42, P43].
& 
Publications focused on security [P42] and maintaining the longevity of software systems [P45].
\\
2020 
&
Publications focus on the cost efficiency of logistics [P53] and maintenance [P55].
& 
No Publications have contributed to environmental sustainability this year. 
& 
Publications focus on trust [P46], employment issues [P47], user-centered design [P49], navigational assistance [P50], collaboration and social good [P52], micro-politics [P53], urban security and tourism [P54], community smells [P56] and user values [P57].
& 
Publications focus on trust and safety in smart software agents [P46], robustness and resistance [P48], maintenance [P51], harms of layered reuse [P53] and technical debt [P55]. 
\\
2021 
&
Only one publication focuses on cost efficiency [P63] this year. 
& 
No publications have contributed to environmental sustainability this year. 
& 
Publications focus on social inclusion [P59, P60, P61, P62, P63, P64, P65, P66, P67, P68], equity [P65, P64, P66],  community engagement [P59, P60, P62], enhancing education [P46], and public health [P68].
& 
Publications focus on the analysis of app reviews [P65],	socio-technical factors [P66], documentation [P67], and image analysis [P68]. 
\\
2022 
&
No publications have contributed to environmental sustainability this year. 
& 
No publications have contributed to environmental sustainability this year. 
& 
Publications in this year mainly focus on inclusivity [P82, P81, P80, P79, P78, P77, P76, P71, P69] and diversity [P76, P77, P85]. 
& 
Publications focus on compatibility [P72], long-term usability [P75], and security [P76]. 
\\
2023 
&
Research has highlighted the economic benefits of incorporating cognitive diversity, emphasizing increased efficiency [P99] and innovation [P99].
& 
No Publications have contributed to environmental sustainability this year. 

& 
Publications mainly focus on inclusivity [P89, P91, P92, P96, P101, P102, P105], gender diversity [P89, P90, P91, P96, P104, P105] and support for marginalized groups [P92].
& 
Publications focus on maintainability [P98], long-term usability [P99, P86], adapting to new requirements [P98], and long-term technical robustness [P99, P86].
\\
2024 
&
Only one study discusses economic sustainability in terms of business continuity of the development environment [P116].
& 
Publications focus on energy Patterns [P107], regulatory compliance for the environment [P110], resource-intensive operations [P116], green architecture tactics [P118], the energy consumption of Large Language Models [P119], and smart local energy [P123]
& 
Publications focus on the impact of humor on developer productivity [P106], imposter syndrome [P115, P120], challenges faced by women [P120], people with ADHD [P111], and people from immigration backgrounds [P108] in the SE workforce, suggestions for early-career developers [P121] and SE research communication [P114]. Others focused on user privacy at run-time [P109], equitable online ecosystem [P110], uncovering algorithmic discrimination [P112], and transparency~[P122].
& 
Publications focus on patterns [P017] and architectural tactics [P118], security and privacy architectures and runtime monitoring [P109], software documentation [P110], algorithmic discrimination [P112], assessment criteria for sustainable SE processes [P116], adaptive UI [P117], prediction accuracy [P119], decision making [P122] and systems of systems [P123]. 
\\
 \hline
\end{tabular} 
\label{tab:sustainability}
\end{table*}

Overall, the trends show an increased focus on approaches addressing the sustainability of the software ecosystem in the context of social issues, followed by a focus on the technical dimension. A lack of publications that address environmental and economic dimensions is observed. 
\section{Discussion}
\label{s:discussion}

\subsection{On Trends and Gaps}
Our analysis of a decade of SEIS publications highlights the main trends and gaps. Overall, the majority of the publications are empirical studies and discuss social aspects. These are categorized as evaluation publications. We observe a scarcity of opinion- and experience-based publications despite these being encouraged in the call for papers. This is an important gap, as it would be valuable for the community to learn from publications that report experience-based perspectives from the viewpoint of, \eg practitioners, researchers, and other disciplines. 

Further, a notable trend is the increasing focus on addressing workplace discrimination in software engineering, with researchers examining its causes, effects, and the importance of fostering inclusive environments. This highlights the need for diversity and anti-discrimination efforts to improve team dynamics and productivity. However, there is a research gap to assess the effectiveness of interventions aimed at mitigating these issues. Studying the social dynamics within the SE workforce could help develop appropriate interventions. Research on the emotional well-being of developers and its impact on productivity also remains limited. Research on the effects of positive discrimination could also add an interesting perspective to the diversity and inclusion issues.

We also observe a trend in ethical concerns that focuses mainly on privacy, discrimination, and misrepresentation. Most of these studies are conducted in the context of a specific community. More research is needed to uncover the challenges and needs of underrepresented groups, directly or indirectly affected by the SE processes and their aftereffects.

Finally, we only found one study [P99] focusing on sharing (\eg as open-source) the knowledge and results for potential reuse. We find this interesting, as the gap between research and practice is still an open problem. We argue that more research should invest in synthesizing results in reusable formats so as to have a greater impact, especially in SE for society.  

\subsection{On Sustainability and Social Impact}

The results of RQ1 show sustainability as the top keyword, and trends also highlight publications discussing sustainability. However, these publications yield a lot of variation in terms of sustainability understanding and representation: while some publications study sustainability in terms of concerns and requirements (both technical and social), others focus on sustainability in communication and governance. 
Also, we see a general misconception of the notion of sustainability, or at least just partial coverage: many publications mention sustainability in some dimension of focus (\eg social or technical), but they neglect, by and large, the sustainability dimension of time, hence disregarding the fact that even positive interventions for, \eg inclusivity are \textit{unsustainable} unless they are accompanied by a durable change in behavior and/or society. More research is needed on how both dimensions of focus and time are combined for true sustainability in SE.

Based on the trends identified in RQ2 and the sustainability mapping in RQ3, our results show that the SEIS track features publications that discuss sustainability for both their \textit{primary} and \textit{enabled} foci.

We see a dominance of the social dimension overall and across different facets of sustainability. The environmental dimension has been observed to be neglected over the years, with a significant decline in recent years. Moreover, only one study [P27] directly addresses the environmental sustainability concerns, while others only achieve it through the enabling impacts \eg employing a technical solution such as design patterns [P107] can lead to energy savings as an enabling effect. 
The same is also true for other dimensions, \eg studying the social impact of UI animations on people with ADHD can enable SE practitioners to develop technical solutions to support their needs. The neglected sustainability dimensions must also be studied for their direct impacts.

Based on our findings, we conclude that a solution aimed at directly addressing one sustainability dimension not only has effects on other sustainability dimensions but also leads to the creation of solutions that are cross-cutting across the four sustainability dimensions of focus. Research considering the impacts of these cross-cutting concerns is essential for achieving sustainability. 

\section{Threats To Validity}
\label{s:threats}
For the discussion of the threats to the validity of our work and related mitigating measures, we referred to the threat categories from Wohlin \etal \cite{wohlin2012}. 

\noindent\textbf{Internal Validity.}
Variability in our findings may be caused by variations in data extraction techniques. To solve this, we created a uniform data extraction procedure and carried out cross-checks across the authors to guarantee precision and consistency.
Naturally, there is a chance of potential bias in the classification and results. However, the authors cross-checked the classification protocol and results to mitigate such bias. We only used keywords from metadata and, in some cases, abstracts and conclusions. There is a change of misrepresentation for the results that solely rely on keyword count per cluster. However, we manually analyzed the keyword groups and updated the topic categories by merging keyword clusters. 

\noindent\textbf{External Validity.}
Since our research is by design limited to the SEIS track, the results cannot be generalized and may be subject to bias based on the specific acceptance criteria of the track itself and/or this specific conference. However, our classification scheme can be applied to other conferences for future comparative analysis of results.  

\noindent\textbf{Construct Validity.}
We systematically followed the pre-defined design of our study for each RQ. Our classification of topics and trends relies on the keywords declared by the authors of the publications under analysis. We analyze the full text of the studies in the top topics to confirm the correctness of the cluster assignment. In our work, none of the publications were removed from or reclassified in another cluster. For the sustainability mapping, we define the sustainability concepts in the background section and use them to classify the publications. The two types of classifications were performed by different authors and, as a mitigation action, were later cross-checked for correctness and consistency.

\noindent\textbf{Conclusion Validity.}
The implications of this research are subject to the researcher's bias. Further, the synthesized results may not capture the complete context of the publications, which can lead to misinterpretation of the results. As a mitigating action, the authors of this study separately performed the synthesis and then cross-checked the results to detect and address potential inconsistencies.
\section{Conclusion}
\label{s:conclusion}

In our review of the ICSE-SEIS track, we analyze the predominant topics, trends, gaps, and sustainability coverage in all SEIS publications so far, from 2015 to 2024. Our findings show that the most common research areas include sustainability, diversity and inclusion, and open-source software. These areas are reflected in trends focusing on sustainability software, gender inequity, inclusivity, ethical concerns in software engineering, and the societal impacts of social media.

We also identify several gaps in the research, particularly concerning publications relevant to low socio-economic groups, the role of emotions in software engineering, discrimination interventions, multi-disciplinary approaches, and the impact of team dynamics on the software engineering process.

Additionally, we evaluate this track through the lens of sustainability. Our results reveal a predominance of social sustainability, while economic and environmental sustainability are discussed less frequently, as well as a noticeable lack in their representation over time.

By reflecting on the evolution of SEIS research over the past decade, our results may help guide fellow researchers to explore SEIS-related new and emerging areas, enabling them to address future societal challenges through software engineering. 

Also, as the gaps in our study emerge from topics addressed by existing, albeit fewer, publications, they are limited to needs that have already been identified. As such, we most certainly miss societal needs where software engineering may help. 
Future work should study the societal needs \eg captured in the governmental agendas and global bodies, and compare them with existing SE results to provide a much more comprehensive research agenda. An analysis of other SE conference tracks that cover the societal impacts could be a useful exercise for our aim. 

\section*{Acknowledgments}
This publication is part of the project SustainableCloud (OCENW.M20.243) of the research programme Open Competition, which is (partly) financed by the Dutch Research Council (NWO), and the LETSGO Project promoted by the Netherlands Enterprise Agency (Rijksdienst voor Ondernemend Nederland). We acknowledge the use of Writeful\footnote{\url{https://www.writefull.com/}} for assistance with language editing in the preparation of this manuscript. 

\bibliographystyle{ieeetr}
\bibliography{main}

\end{document}